# A Cosmic Ray Muon Spectrometer Using Pressurized Gaseous Cherenkov Radiators

Junghyun Bae, Stylianos Chatzidakis

*Abstract*—In this work, we propose a new approach to cosmic ray muon momentum measurement using multiple pressurized gaseous Cherenkov radiators. Knowledge of cosmic ray muon momentum has the potential to significantly improve and expand the use of a variety of recently developed muon-based radiographic techniques. However, existing muon tomography systems rely only on muon tracking and have no momentum measurement capabilities which reduces the image resolution and requires longer measurement times. A fieldable cosmic ray muon spectrometer with momentum measurement capabilities for use in muon scattering tomography is currently missing. We address this challenge by optimally varying the pressure of multiple gaseous Cherenkov radiators and identifying the radiators that are triggered by muons that have momentum higher than the Cherenkov threshold momentum. We evaluate the proposed concept through Geant4 simulations and demonstrate that the cosmic ray muon momentum spectrum can be reconstructed with sufficient accuracy and resolution for two scenarios: (i) a perfect Cherenkov muon spectrometer and (ii) a practical spectrometer where noise is introduced in the form of scintillation and transition radiation photons. To quantify the accuracy of spectrometer, the concept of true and false classifications are introduced. The fraction of true classification is investigated for each momentum level in a practical radiator. The average classification rate for momentum range of 0.2 to 7.0 GeV/c with uncertainty of 1 GeV/c is approximately 85%.

## I. INTRODUCTION

COSMIC ray muon radiography, or muography, is a promising non-destructive technique that is often utilized to monitor or image the contents of dense or well shielded objects, typically not feasible with conventional radiography techniques. Although a muography has been used with various levels of success in imaging of spent nuclear fuel casks and nuclear reactors [1]–[5], nuclear safeguard and homeland security [6]–[9], and geotomography [10]–[13], it suffers from long measurement times limiting its widespread applicability. The benefits of measuring muon momentum coupled with existing tomography systems have been discussed [14], [15], however, it is still very challenging to deploy existing muon spectrometers in the field due to size and cost limitations. Recent efforts develop techniques to estimate muon momentum for use in muon scattering tomography [16], [17] and to improve the accuracy of cosmic muon flux estimation [18]–[20] look promising, however, a portable prototype has yet to be developed. In this work, we present a concept for measuring muon momentum using multiple gaseous Cherenkov radiators. By varying the pressure of multiple gas Cherenkov radiators, a set of muon momentum threshold levels can be selected that are triggered only when the incoming muon momentum exceeds that level. As a result, depending on the incoming muon momentum, none to all Cherenkov radiators can be triggered. By analyzing the signals from each radiator, we can estimate the actual muon momentum. The primary benefits of such a concept is that is can be compact and portable enough so that it can be developed in the filed separately or in combination with existing muon tomography systems.

## II. MUON MOMENTUM MEASUREMENT USING CHERENKOV RADIATION

When a muon travels through a transparent medium at a speed greater than the speed of light in that medium, it emits Cherenkov radiation. The threshold muon momentum, $p_{th}$, needed to induce Cherenkov radiation is:

$$p_{th}c = \frac{m_\mu c^2}{\sqrt{n^2-1}} \quad (1)$$

where $n$ is the refractive index, $m_\mu c^2$ is the muon rest mass, and $c$ is the speed of light in vacuum. For gaseous media, the refractive index is a function of pressure and temperature and it can be estimated using the Lorentz-Lorenz equation [21]:

$$n \approx \sqrt{1 + \frac{3Ap}{RT}} \quad (2)$$

where $A$ is the molecular refractivity in m$^3$/mol, $R$ is the universal gas constant, $p$ is pressure and $T$ is temperature. From (2), the refractive index, $n$, can be varied by pressurizing a gaseous radiator at a constant temperature. The gas properties of the selected Cherenkov radiators are summarized in Table 1. The conceptual design of the proposed muon spectrometer is shown in Fig. 1. Each numbered box represents a gas radiator with various muon momentum threshold levels. Given that a muon with an energy of 3.1 GeV traverses all radiators, optical signals due to the Cherenkov radiation emission is only

J. Bae and S. Chatzidakis are with the School of Nuclear Engineering, Purdue University, West Lafayette, IN 47906 USA (corresponding author e-mail: bae43@purdue.edu).

TABLE I. GAS PROPERTIES OF THE SELECTED CHERENKOV RADIATORS ($CO_2$, $C_4F_{10}$, AND AIR) [22]–[28]

|  | $CO_2$ | $C_4F_{10}$ | Air |
|---|---|---|---|
| Molecular Weight [g/mol] | 44.01 | 238.03 | 28.96 |
| Molecular Polarizability [$10^{-30}$ m$^3$] | 2.59 ± 0.01 | 8.44 ± 0.12 | 21.18 ± 0.91 |
| Refractive Index | 1.00045 | 1.0014 | 1.000273 |
| Scintillation Efficiency [#/MeV] | 5.09 ± 0.28 | 3.1 ± 1.6 | 25.46 ± 0.43 |

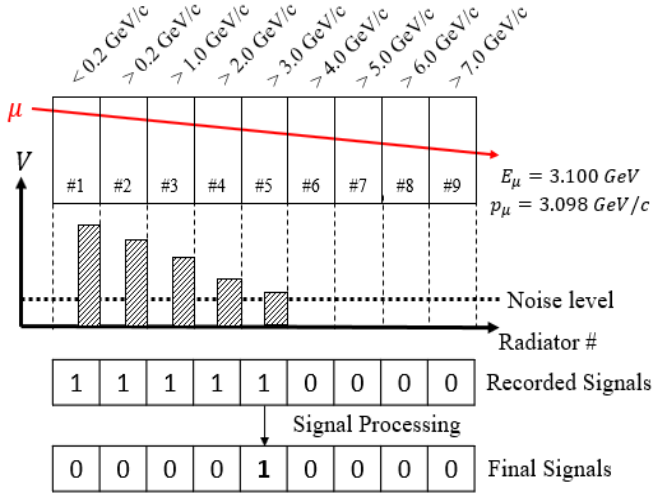

Fig. 1. Overview of signal processing of multiple gaseous Cherenkov radiators for a 3.1 GeV muon. The final signal correctly identifies the actual muon momentum which is 3.0 – 4.0 GeV/c.

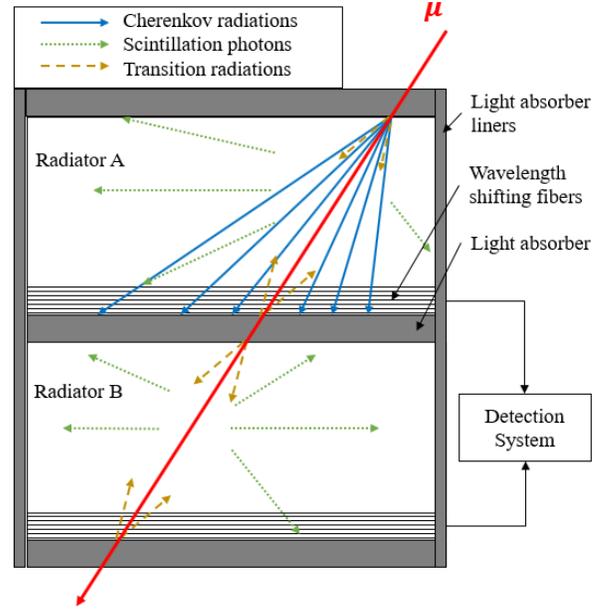

Fig. 2. Characteristic photon emissions by Cherenkov, scintillation, and transition radiation in radiator A ($p_\mu > p_{th}$) and radiator B ($p_\mu < p_{th}$) when a single muon passes through both radiators. All photon signals are transmitted to the detector system via wavelength shifting fibers.

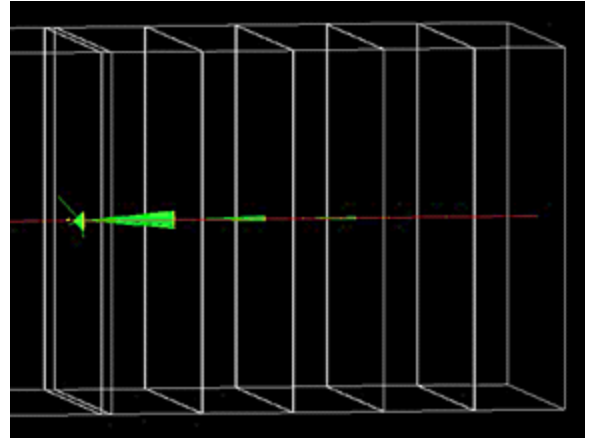

Fig. 3. The emission of Cherenkov radiation (green) in a perfect muon spectrometer (scintillation and transition radiation are excluded) using Geant4 simulation. Cherenkov radiation is only observed in the first four radiators because the incoming muon (red) momentum exceeds their threshold momentum levels [28].

expected in radiators up to #5 (> 3.0 GeV/c) because the actual muon momentum is greater than 3.0 GeV/c. Furthermore, amplitudes of the expected Cherenkov photon signals decrease as radiator number increases since the refractive index becomes smaller for higher threshold momentum levels (lower pressure). After processing the recorded binary signals, the final signals correctly identify the actual momentum range, 3.0 to 4.0 GeV/c.

III. RESULTS

To evaluate and validate the feasibility of our proposed muon spectrometer, we performed detailed Monte Carlo simulations using Geant4. The cosmic muon momentum spectrum at sea level was simulated using a MATLAB open source named "Muon_generator_v3 [29]" which is developed based on the semi-empirical models. We then categorized the incoming muon momentum into seven momentum levels from 0.2 to 7.0 GeV/c with an interval of 1.0 GeV/c. In a perfect radiator, signals are exclusively generated by Cherenkov radiation. In a practical radiator, on the other hand, scintillation photons and transition radiation are included besides Cherenkov radiation. The characteristic photon emissions by Cherenkov, scintillation, and transition radiation in radiator A ($p_\mu > p_{th}$) and radiator B ($p_\mu < p_{th}$) are shown in Fig. 2. The visualized Geant4 simulation result for the Cherenkov radiation emission is shown in Fig. 3. Radiators that have a threshold momentum level lower than the actual muon momentum emits Cherenkov radiation. The results for the cosmic ray muon spectrum (a), and reconstruction spectra using a (b) perfect muon spectrometer and (c) practical muon spectrometer are shown in Fig. 4. A perfect muon spectrometer accordingly categorizes the incoming muon momentum in the proper ranges with a 100% accuracy due to the absence of noise. On the other hand, a practical muon spectrometer sometimes incorrectly categorizes the actual muon momentum and it is represented as the false classification rate. Our proposed muon spectrometer classified the wide range of muon momenta with a high accuracy (~85%) and successfully reconstructed the cosmic ray muon spectrum (Fig. 4-(c)).

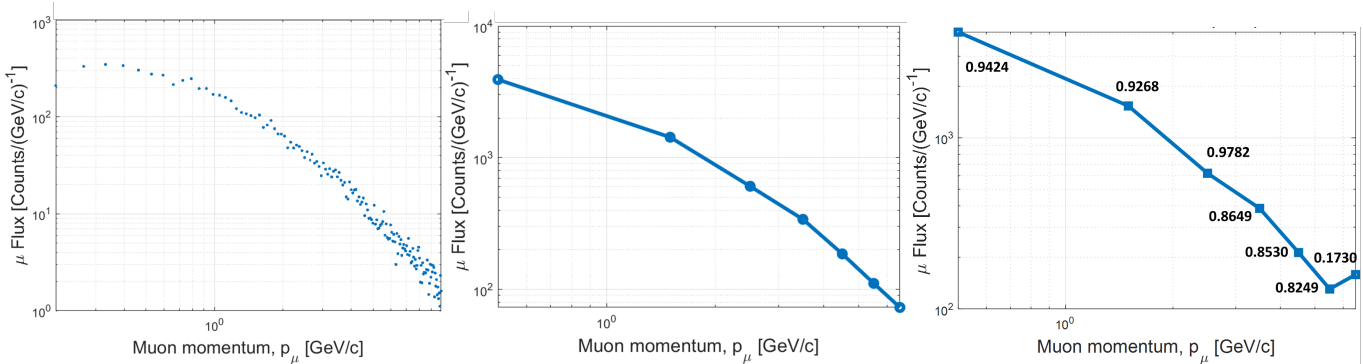

Fig. 4. (a) The actual cosmic muon spectrum using the semi-empirical models [29] and reconstructed spectra using (b) a perfect and (c) a practical muon spectrometer. The true classification rates for momentum levels are also shown in (c).

## IV. Conclusion

Knowledge of the cosmic muon momentum can play a significant role in muon scattering tomography applications. In this paper, we propose a new concept for muon momentum measurement using pressurized gaseous Cherenkov radiators. We showed that the use of multiple pressurized gas Cherenkov radiators has the potential to successfully measure the incoming cosmic muon momentum with a measurement resolution of 1.0 GeV/c within the range between 0.2 and 7.0 GeV/c. To demonstrate the feasibility of our proposed spectrometer prototype, we performed the Geant4 simulations. Our results show that the reconstructed spectrum is in good agreement with the actual cosmic ray muon spectrum and the average measurement true classification rate is approximately 85%.

## Acknowledgment

This research is being performed using funding from the Purdue Research Foundation.